\documentclass[useAMS,usenatbib]{mn2e}
\usepackage{epsfig}

\def\gtrsim{\lower 2pt \hbox{$\, \buildrel {\scriptstyle >}\over
{\scriptstyle \sim}\,$}}
\def\lesssim{\lower 2pt \hbox{$\, \buildrel {\scriptstyle <}\over
{\scriptstyle \sim}\,$}}

\def\chandra{{\sl Chandra}}
\def\spitzer{{\sl Spitzer}}
\def\hst{{\sl HST}}
\def\pa{P{$\alpha$}}
\newcommand{\as}{$^{\prime\prime}~$}

\pdfoutput=1

\title[HST/NICMOS Paschen-$\alpha$ Survey of the Galactic Center: Overview]{HST/NICMOS Paschen-$\alpha$ Survey of the Galactic Center: Overview}
\author[]{Q. D. Wang$^{1}$\thanks{E-mail:
wqd@astro.umass.edu}, H. Dong$^{1}$, 
A. Cotera,$^{2}$ S. Stolovy,$^{3}$ M. Morris,$^{4}$ C. C. Lang,$^{5}$ 
\newauthor M. P. Muno,$^{6}$ 
 G. Schneider,$^{7}$ and D. Calzetti$^{1}$\\
$^{1}$Department of Astronomy, University of Massachusetts, 
  Amherst, MA 01003, USA\\
$^{2}$SETI Institute, 515 North Whisman Road, Mountain View, CA 94043, USA\\
$^{3}$Spitzer Science Center, California Institute of Technology, Mail Code 220-6, 1200 East California Boulevard, Pasadena, CA 91125, USA\\
$^{4}$Department of Physics and Astronomy, University of California, Los Angeles, CA 90095, USA\\
$^{5}$Department of Physics and Astronomy, University of Iowa, Iowa City, IA 52245, USA\\
$^{6}$Space Radiation Laboratory, California Institute of Technology, Pasadena, CA 91125, USA\\
$^{7}$Steward Observatory, The University of Arizona, Tucson, AZ 85721, USA}

\begin{document}

\date{Accepted 2009 November 2;  Received 2009 October 29; in original form 2009 September 22}

\pagerange{\pageref{firstpage}--\pageref{lastpage}} \pubyear{2009}

\maketitle

\label{firstpage}

\begin{abstract}
We have recently carried out the first wide-field hydrogen Paschen-$\alpha$ 
line imaging survey of the Galactic Center (GC), using the NICMOS instrument 
aboard the Hubble Space Telescope. The survey maps out a region of 
$2253 {\rm~pc^2}$ ($416 {\rm~arcmin^2}$) around the central supermassive 
black hole (Sgr A*) in the 1.87 and 1.90 \micron\ narrow bands 
with a spatial resolution of $\sim 0.01$ pc 
(0\farcs2 FWHM) at a distance of 8 kpc. Here we present an overview of 
the observations, data reduction, preliminary results, and potential scientific implications, 
as well as a description of the rationale and design of the survey. 
We have produced mosaic maps of the Paschen-$\alpha$ line and continuum 
emission, giving an unprecedentedly high resolution and high sensitivity 
panoramic 
view of stars and photo-ionized gas in the nuclear environment of the Galaxy. 
We detect a significant number of previously undetected stars with 
Paschen-$\alpha$ in emission.
They are most likely massive stars with strong winds, as confirmed
by our initial follow-up spectroscopic observations. 
About half of the newly detected massive stars 
are found outside the known clusters (Arches, 
Quintuplet, and Central). 
%Part of this new population of massive stars may be dynamically ejected from the clusters, while many of the others are apparently embedded in distinct HII regions, probably representing massive stellar clusters/groups in formation.
Many previously known diffuse thermal 
features are now resolved into arrays of intriguingly fine 
linear filaments indicating a profound role 
of magnetic fields in sculpting the gas. The
bright spiral-like Paschen-$\alpha$ emission around Sgr A* is seen to be
well confined within the known dusty torus. In the directions roughly
perpendicular to it, we further detect faint, diffuse Paschen-$\alpha$ 
emission features, which, like earlier radio images, suggest an 
outflow from the structure. In addition, we detect various compact 
Paschen-$\alpha$ nebulae, probably 
tracing the accretion and/or ejection of 
stars at various evolutionary stages. Multi-wavelength comparisons
together with follow-up observations are helping us to address such 
questions as where 
and how massive stars form, how stellar clusters are disrupted, how
massive stars shape and heat the surrounding medium, how various phases of this
medium are interspersed, and how the 
supermassive black hole interacts with its environment. 
\end{abstract}


\section{Introduction}
\label{s:intro}

The Galactic center (GC) is a unique laboratory for a detailed study 
of star formation and its impact on the nuclear environment of galaxies.
Because of its proximity, the GC can be imaged at resolutions 
more than 100 times better than the nearest galaxies like our own
(e.g., the Andromeda galaxy). This unmatched high-resolution capability
is particularly important to the study of the formation and destruction 
processes of stellar clusters as well as the overall population of massive
stars in such an environment. Indeed, the GC is known to host three 
very massive clusters~\citep[the Arches and Quintuplet clusters, as well as 
the central parsec cluster surrounding Sgr A$^*$; ages $\sim 2-7 \times 
10^6$ yrs, masses $\sim 3\times10^{4} M_\odot$ each; e.g.,][]{cot96,ser98,fig99,fig02,gen03}, which are responsible for 
about 5\% of the total Lyman continuum flux for the Galaxy, in less
than 0.01\% of its volume. 
Massive stars are known to exist outside the clusters
~\citep{cot06}, however, locating these important stars has been
sporadic and incomplete~\citep[e.g. ][]{mun06a,mau07,mau09}. These isolated massive stars may have formed individually or in small groups, although the extreme conditions in the GC make this mode of star formation unusually difficult.   They may also 
have been spun out of the clusters due to their internal
dynamics. This process is expected to be accelerated by the strong
external tidal force in the GC~\citep{kim99}.

\begin{figure*} %[tbh]
 \centerline{
  \includegraphics[width=1.1\textwidth,angle=0]{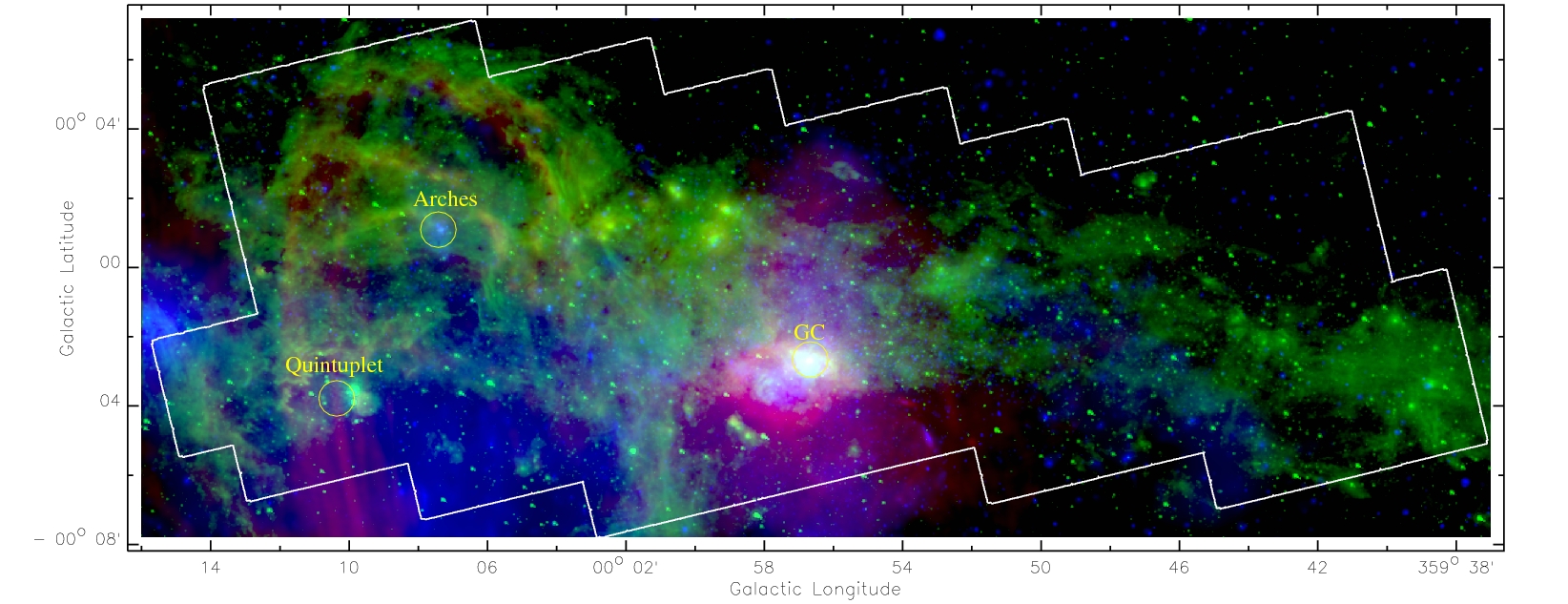}
 } 
\caption{%\footnotesize 
A multi-wavelength montage of the GC: the VLA 20 cm continuum~\citep[red; ][]{yus84}, \spitzer\ 8 $\mu$m~\citep[green; ][]{stolo06,are08}, and \chandra\ ACIS-I 1-9 keV~\citep[blue; ][]{wan02,mun09}. The three known massive young
clusters are shown, as well as the outer border of the \pa\ survey.
} \label{f:fig1}
\end{figure*}
\bigskip

The star formation process depends critically on the interplay of 
massive stars with the interstellar medium (ISM). Massive stars 
release large amounts of ultraviolet radiation. They also produce strong
stellar winds throughout their short lives
and die in supernova explosions. Such energetic stellar 
output dramatically shapes the morphology of the ISM, as is  
revealed in existing radio, mid-infrared, and X-ray maps
~\citep[Fig.~\ref{f:fig1}; e.g., ][]{yus84,yus02,wan02,wan06,mun09}. 
Perhaps the best-known example of
this is the Radio Arc region which surrounds the Arches and Quintuplet
clusters: a collection of arc-like thermal filaments,
synchrotron-bright non-thermal linear filaments, and diffuse, 
hot gas. Some of the structures within this region, the Sickle HII region 
ionization front and photoevaporated pillars in particular, suggest that 
gas is being collected and compressed~\citep[][Cotera et al. in preparation]{cot06}, and is morphologically resembles  known regions 
outside the GC where such processes are believed to have led to the 
formation of a new generation of stars (e.g. M16; \citet{smi05}).
The energy injection from massive stars and possibly from the 
central supermassive black hole (SMBH), that is coincident with the radio source 
Sgr A*, may also strongly affects the thermal and/or dynamical properties of the 
ISM. Molecular gas in the GC, for example, has unusually high
turbulent velocities and temperatures compared to clouds in other parts
of the Galaxy~\citep[e.g., ][]{mor96}. This may bias the
stellar initial mass function toward heavier stars and favor the
production of massive star clusters~\citep[][ and
references therein]{mor93,wan06}. 

To constrain the dynamical process of the known clusters, the 
overall population of massive stars, and their formation modes,
we sought to obtain a uniform survey of massive stars and their interplay with
the ISM across the central 90 parsecs. Ionized warm gas provides the ideal tracer 
of massive stars and their interplay with their surroundings. 
Although previous radio continuum observations are free of
extinction and trace both thermal and non-thermal emission,
the scale-dependence of radio 
interferometers makes it difficult to accurately measure flux densities on 
arcsecond scales. In addition, fine features (e.g., ionization fronts) are often
found in the midst of diffuse non-thermal emission (e.g., the Radio Arc), 
making decomposition of thermal and non-thermal components
challenging.  Radio recombination line observations typically
have a substantially lower signal-to-noise and spatial
resolution than radio continuum data. In the near-IR, the
2.16\micron\ Br$\gamma$ line is accessible from the ground (no suitable 
filter is available for existing and planned space-based platforms), but 
has also not proved effective for
large scale mapping of warm ionized gas. To our knowledge, no large-scale
Br$\gamma$ line survey of the GC has ever been published. The difficulty
lies in the lack of the combination of high spatial resolution and 
stable point spread function (PSF)
in a ground-based survey. Observations with adaptive optics (AO) can
have superb resolution, but only over small fields of
view (typically $15\arcsec - 40\arcsec$). The AO PSF also changes
strongly, both spatially and with time, and has extended PSF wings,
%(containing $\sim 70\%$ of the power, compared to 13\% for NICMOS/NIC 3), 
making the PSF subtraction extremely difficult.  In addition, photometric
conditions rarely, if ever, persist for the time required for such a survey, 
hampering the study of large-scale features. 

We have carried out the first large-scale, high-resolution, near-IR
survey of the GC, using {\it HST} NICMOS. The primary objective of
the survey is to map out the hydrogen Paschen-$\alpha$ (P$\alpha$ for short) 
line (wavelength 1.876 $\micron$) in a field of $\sim 39^\prime \times 15^\prime$ 
around Sgr A$^*$~\citep[corresponding to 90 pc $\times$ 35 pc at the GC distance of
8~kpc; ][
Fig.~\ref{f:fig1}]{ghe08}. This field is known to be rich in 
clustered massive star formation (i.e., including the three known clusters). 
NICMOS observations of the 
P$\alpha$ emission were taken previously only for a few
discrete regions, centered on the three known massive clusters
 (see \S~\ref{ss:comparion} for further discussion). Our survey field also includes 
much unexplored territory. In
particular, the region between Sgr A and Sgr C along the Galactic
plane (to the right in Fig.~\ref{f:fig1}) shows a number of compact {\it Spitzer} IRAC
sources (several with radio counterparts) likely related to some of the 
earliest stages of massive star
activities. This field selection also facilitates an unbiased
study of the apparent asymmetry between the positive and negative
Galactic longitude sides relative to Sgr A$^*$. 
%This asymmetry is somewhat puzzling, because the orbital periods of matter in this field are short ($\lesssim$2 Myr), shorter than  the ages of the star clusters.

\begin{figure*} %[tbh]
 \centerline{
 \includegraphics[width=1.15\textwidth,angle=0]{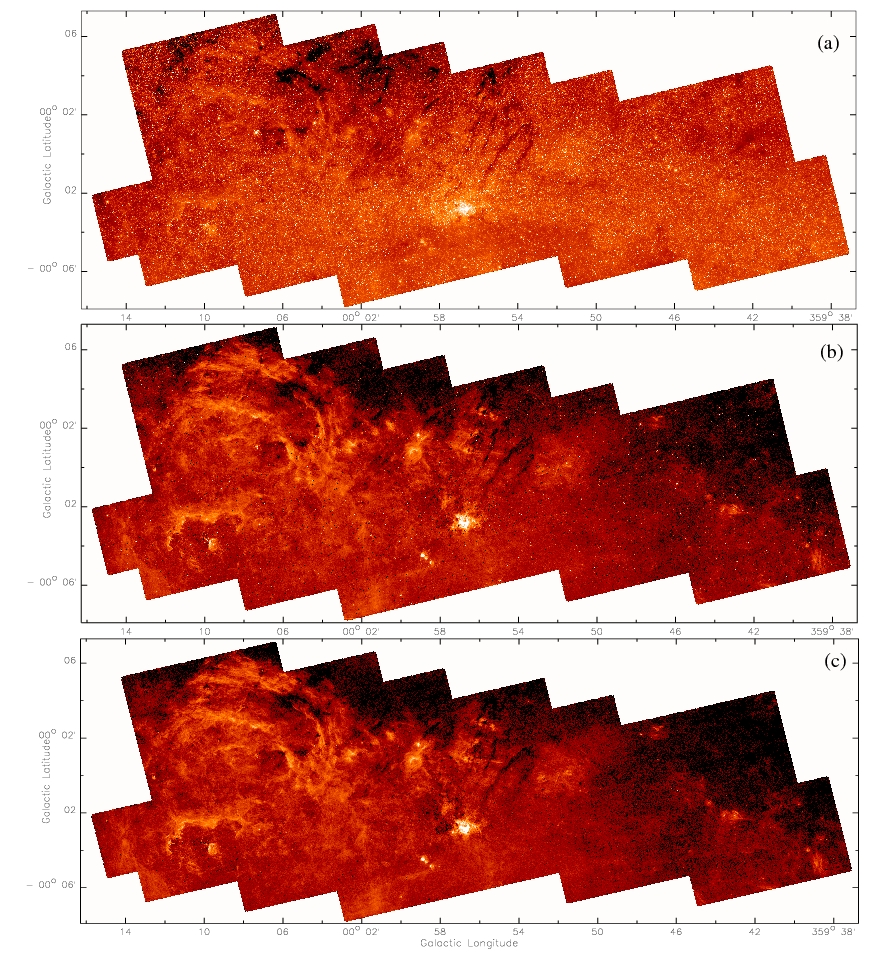}
 } 
\caption{%\footnotesize 
NICMOS maps from our survey: (a) calibrated F187N mosaic,
(b) continuum-subtracted F187N map, and (c) continuum-subtracted and 
point source removed F187N  (diffuse \pa) map. These maps are
presented with a bin size of 0\farcs4. The intensity is logarithmically 
scaled from of 0.01 ${\rm~mJy~arcsec^{-2}}$ to 10 ${\rm~mJy~arcsec^{-2}}$ 
in (a) and to 3 ${\rm~mJy~arcsec^{-2}}$ in (b) and (c). 
%\pa-emitting sources are marked in green circles.
} \label{f:fig2}
\end{figure*}
\bigskip

Our \hst\ NICMOS observations overcome 
all of the potential problems of a ground-based survey.  
The wide NIC3 camera on this instrument provides a resolution of $\sim0\farcs2$ with a stable
PSF. This capability is important for detecting and cleanly removing
point-like sources as well as resolving fine structures of extended
features. Both the stable PSF and the absence of atmosphere for 
our \hst\ survey means that the entire survey is photometrically
accurate and consistent. 
The extra magnitude of extinction along the sight-line 
to the GC at 1.87\micron\ (as compared to 2.16 \micron) is more than compensated 
for by an intrinsic P$\alpha$ line intensity 12 
times greater than the Br$\gamma$ line~\citep{hum87}. In addition, the background 
at P$\alpha$ with NICMOS 
is a factor of $\sim 800$ lower than the sky background at Br$\gamma$
as observed from the ground. This low background of \hst\
NICMOS also allows us to detect faint point-like sources
and extended features, particularly over our large mosaicked field.

\section{Observations and data reduction}
\label{s:ob}
Our survey was completed in 144 HST orbits between February 22 and June 5,
2008. All of these orbits had an identical design, 
each consisting of $2\times 2$ NIC3 pointing positions for 
both F187N (on-line) and F190N (off-line) filters.
These filters have a 1\% bandpass centered at the wavelengths 1.875$\micron$ and 
1.900 $\micron$. NIC3 is a 256$\times$ 256 HgCdTe array detector with a pixel size of $\sim 0\farcs2031 \times 0\farcs2026$ in the x and y directions. Adjacent
positions and orbits were designed to have 3\as\ overlaps in the vertical 
direction of the NIC3 array coordinates, partly to allow for the bottom 
15 pixels that do not provide useful data due to afocal
vignetting of that portion of the array. This is in addition to the
overlaps resulting from the 4-point dithers at each of the 
four positions of an orbit. Each dither exposure was taken 
in the MULTIACCUM readout mode (SAMP-SEQ=16; NSAMP=9) 
for photometry over a large flux range and for effective 
identification and removal 
of cosmic-rays and saturated pixels. The 4-point dithers
were in an inclined square wave pattern (i.e., the shifts were
 in the directions of 67$^\circ$.5 relative to the x-axis of the detector 
coordinates); each dither step produced a shift of
2\farcs33 and 5\farcs63 pixels in 
the x and y directions. This dither pattern enabled both a 
sub-pixel sampling of the undersampled NIC3 PSF
%in near orthogonal (67$^\\circ$.5$) array coordinates 
and a consistent overlap between orbits.
% with a 30\farcs5 overlap between adjacent positions??? 
%These shifts, together with the dithering pattern, further allow for effective removal of relative biases among individual quadrants in each of the four dithers. 
Each orbit effectively covered an area of roughly
102$^{\prime\prime}\times98^{\prime\prime}$ (2.8 arcmin$^2$).  
With this strategy and a single telescope orientation, the survey
mapped out a field of 416 arcmin$^2$ 
(as outlined in Fig.~\ref{f:fig1}),
and with a uniform 192 sec exposure per filter. The entire science data 
set includes
144 (orbits) $\times$
4 (pointing positions) $\times$ 4 (dithers)  $\times$ 2 (filters) $= 4604$ images.

In addition, between target
visibility periods (during Earth occultations) after each of the 144 orbits, 
16 dark frames were obtained with the detector
clocked in a manner identical to the science exposures. These
inter-orbit raw dark exposures were assembled to form a ``superdark'' frame
used in our image calibration.

\begin{figure} %[tbh]
%\vspace{-0.5pc}
%\vspace*{0.1pc}
 \includegraphics[width=0.485\textwidth,angle=0]{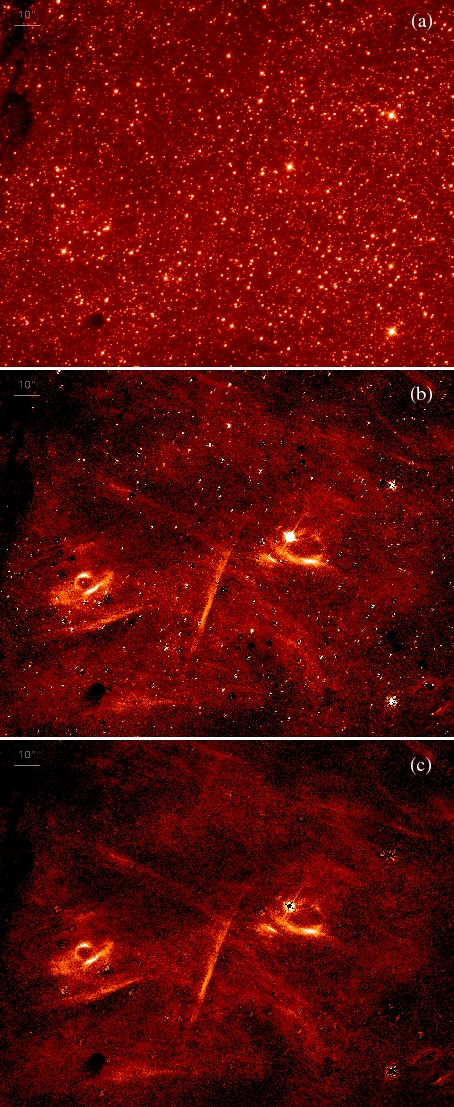}
 %\centerline{\epsfig{figure=fig3.eps,width=0.485\textwidth,angle=0}}
\caption{%\footnotesize 
A full-resolution (0\farcs1 pixel) close-up of a newly discovered P$\alpha$-emitting complex 
(G359.866+0.002), about 5$^\prime$ northwest of Sgr A*: (a) F187N image, (b) continuum-subtracted F187N image,
and (c) point source removed (diffuse \pa) image.
The intensity is logarithmically scaled over the range
of 0.1-30 ${\rm~mJy~arcsec^{-2}}$ in (a) and 0.1-1 
${\rm~mJy~arcsec^{-2}}$ in (b) and (c).}
\label{f:fig3}
\end{figure}

The data reduction and analysis, based on a combination
of several newly developed tools and the standard routines 
implemented in the software IRAF/STSDAS, will be detailed in~\citet{don09}. Briefly, in addition to the standard DC offset correction, we use 
a global fitting method to determine the relative instrumental background 
offsets among all NIC3 array quadrants, pointing positions, and orbits, 
based on multiple overlapping fields.
A similar method is also used to optimize the correction for the relative
astrometry offsets among orbits. The absolute astrometry is corrected 
in respect to radio counterparts of the near-IR sources in the field, 
reaching an accuracy better than $0.1''$. The source detection is based primarily on the {\sl Starfinder} software~\citep{dio00}.

The procedure and product of the data reduction are illustrated in Fig.~\ref{f:fig2}
for global views and Fig.~\ref{f:fig3} for full-resolution close-ups.
To estimate the continuum contribution
in the F187N band (Figs.~\ref{f:fig2}a and \ref{f:fig3}a), 
we adaptively construct a mean F187N to F190N intensity
ratio map, with the value of each pixel given as the median ratio of the 
nearest 100 sources, most of which are assumed to be located at the GC. 
However, this assumption does not hold for regions with exceptionally high 
extinction, as judged from a low surface number density of detected sources 
(i.e., $\lesssim 0.3$ sources~arcsec$^{-2}$).
For such regions, we calculate the ratio values
from the extinction map obtained from {\sl Spitzer} observations
~\citep[][ for details]{sch09,don09}. The 
continuum-subtracted map (Figs.~\ref{f:fig2}b and \ref{f:fig3}b) gives a crude look
of the potential \pa-emitting sources as well as the diffuse emission.
However, one cannot judge whether or not a peak (e.g., in Fig.~\ref{f:fig2}b 
or \ref{f:fig3}b) is a \pa\ emission source based only on its 
apparent 
excess above the background; such a peak can easily be caused 
by the statistical flux fluctuation of a subtracted source, especially a 
bright one. In fact, all peaks in \ref{f:fig3}b, except for the brightest one, 
 are below our current \pa\ source detection  threshold. 
We identify \pa\ source candidates based on their integrated 
continuum-subtracted fluxes at a statistical 
confidence  $\gtrsim 5\sigma$, where $\sigma$ accounts for all 
known measurement errors~\citep{don09}.
To produce a {\sl diffuse} \pa\ map (Figs.~\ref{f:fig2}c and \ref{f:fig3}c), we  use the actual F187N to F190N
flux ratio of a detected source for its affected pixels
(defined to have its PSF-predicted intensity contribution 
$\gtrsim 5\sigma$ above the local background in the F190N map). 
We further 
remove any uniformly-distributed foreground and residual 
instrument background in a map by shifting its intensity to zero, averaged 
over a few darkest regions (see \S~\ref{s:results}).

\section{Preliminary results}
\label{s:results}

Fig.~\ref{f:fig2} shows a panoramic view of our survey data.
Clearly, the distribution of both the stellar light and the 
diffuse emission is highly 
inhomogeneous, partly due to the variations in foreground extinction, which
 varies strongly from one region to another and on different
spatial scales. 
%On average, the extinction toward the GC is known to be around $A_K \sim 3$ (4.1 and 4.0 in the NICMOS bands, F187N and F190N; and 1.3 in the IRAC 8\micron\ band). Based on the F187N/F190N intensity ratios of detected stars, we have constructed an extinction map, which is used for correcting the extinction in our survey (Dong et al. 2009). 
The most outstanding extinction 
features in Fig.~\ref{f:fig2}
are the distinct dark filaments (tendrils) in the field just above Sgr A,
which are also noticeable in published K-band images~\citep[e.g., ][]{phi99}. The darkness of these filaments
indicates that they are foreground thick dusty clouds silhouetted against the 
IR-luminous nuclear region of the Galaxy. The filaments appear to be 
structured down to arcsecond scales [or $\sim 10^{16}(d/1 {\rm kpc})$ 
cm, where $d$ is the distance to such a filament; see the full resolution
close-ups in Fig.~\ref{f:fig3}]. With our data, we 
can distinguish between extinction and
lack of ionized emission by comparing the
\pa\ and F190N images -- if a region is dark 
in both images, we can say with confidence that the dark features indicate regions with extinctions Av $> 40-50$.

\subsection{Extended \pa\ Emission}

Fig.~\ref{f:fig2}c shows
extended \pa-emitting complexes across much of the field. Strong and 
distinctive emission features are seen on angular scales from sub-arcsecond 
to about 10$^\prime$, with surface brightness ranging from as high as $\sim 10
{\rm~mJy~(milliJansky)~arcsec^{-2}}$ down to a 3$\sigma$ level of $\sim 0.05 {\rm~mJy~arcsec^{-2}}$
for the image with the 0\farcs4 $\times$ 0\farcs4 pixel. The
rms noise of a map depends on its pixel size and is $\sim 0.06 {\rm~mJy~arcsec^{-2}}$ 
for the full resolution image with a pixel size of 0\farcs1 $\times$ 0\farcs1,
accounting for both statistical and systematic fluctuations (photon counting,
artifacts from stellar continuum subtraction, etc.).
This noise is estimated from selected regions in the \pa\ map 
~\citep[e.g., at the upper right corner of the map; see ][ for details]{don09} 
that show little \pa\ emission but are bright in the F190N continuum image
(hence the rms contribution from clumpy structures in
the \pa\ emission or extinction should be minimal). The overall distribution
of the \pa\ emission is highly lopsided toward the positive Galactic 
longitude side of the GC, similar to what is seen in radio and mid-IR
(Fig.~\ref{f:fig1}).

\begin{figure*} %[t]
%\vspace{-0.5pc}
%\vspace*{0.1pc}
 \includegraphics[width=1.\textwidth,angle=0]{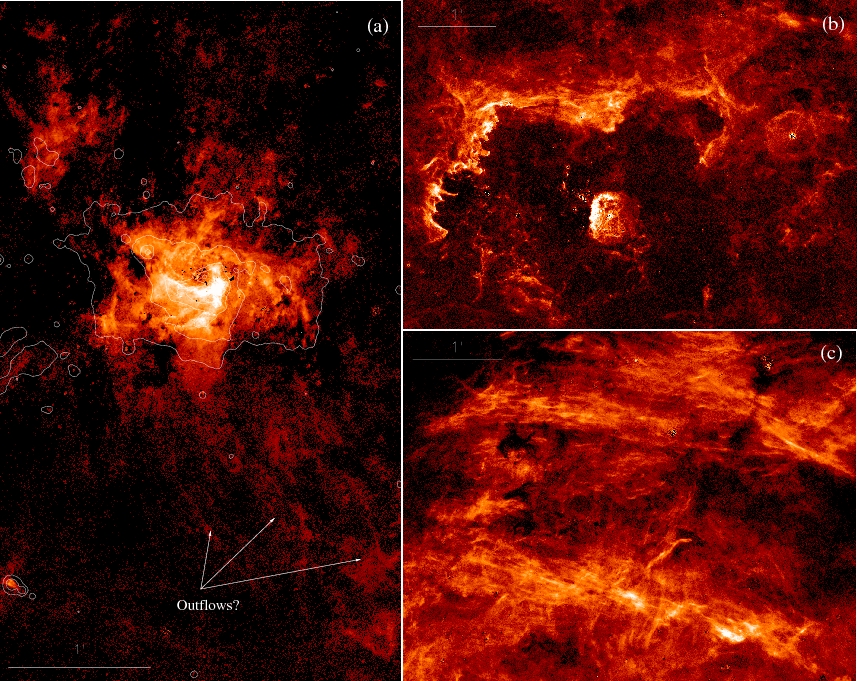}
 %\centerline{\epsfig{figure=fig4.eps,width=1\textwidth,angle=0}}
\caption{%\footnotesize 
Close-ups of the P$\alpha$-emitting features:
(a) the central region around Sgr A*, with overlaid IRAC 8\micron\ intensity
contours at 1,3,10, and 30  $\times 10^3 {\rm~mJy~sr^{-2}}$; 
(b) the Sickle nebula (G0.18-0.04; e.g., \citet[][]{lan97}); (c) 
thermal Arched filaments~\citep[e.g., ][]{lan01}. All are projected in the 
Galactic coordinates. }
\label{f:fig4}
\end{figure*}

The diffuse  \pa\ emission is greatly enhanced around the three 
known clusters responsible for the ionization of the inner rims of 
surrounding gas. 
The emission around the GC cluster is largely confined within a small 
spiral-like nuclear structure, the \pa\ emission of which was first
detected by~\citep{sco03}. Our new 
survey now enables us to trace the faint emission 
features to large distances, which extend 
vertically below and above this structure (Fig.~\ref{f:fig4}a). 
These features may represent outflows from the central few parsecs of the galaxy.
The bright \pa\ emission itself indicates the ionized inner portion of a dusty 
circumnuclear disk
around Sgr A$^*$. This disk appears more extended, as traced by
the 8 \micron\ emission (Fig.~\ref{f:fig4}a). 
Detailed analysis/modeling of these structures is needed 
to further investigate the physical state of the disk and the energetics
of the GC cluster and/or the SMBH, which is  currently in a radiatively 
inefficient accretion state.

Fig.~\ref{f:fig4}b presents
a close-up of the Sickle HII region, revealing fingers of ionized gas 
resembling the ``Pillars of Creation" in M16~\citep{cot06}. These fingers are apparently
illuminated by ionizing photons from hot stars in the Quintuplet cluster~\citep[][Cotera et al. in preparation]{cot06, sim97}.
The \pa-emitting ``Arches'' around the Arches cluster shows an amazing array
of fine linear structures
(Fig.~\ref{f:fig4}c), which are preferentially oriented in the same 
directions as magnetic field vectors measured in infrared polarization
~\citep{chu03,nis09}. Similar 
organized narrow linear \pa-emitting
features are also abundant in some of the more compact HII regions
seen in Fig.~\ref{f:fig2}c. These structures may 
indicate a critical role that locally strong 
magnetic fields play in shaping the morphology and dynamics of the 
ionized gas~\citep[e.g., ][]{nis09}

\begin{figure*} %[t]
%\vspace{-0.5pc}
%\vspace*{0.1pc}
  \includegraphics[width=1.\textwidth,angle=0]{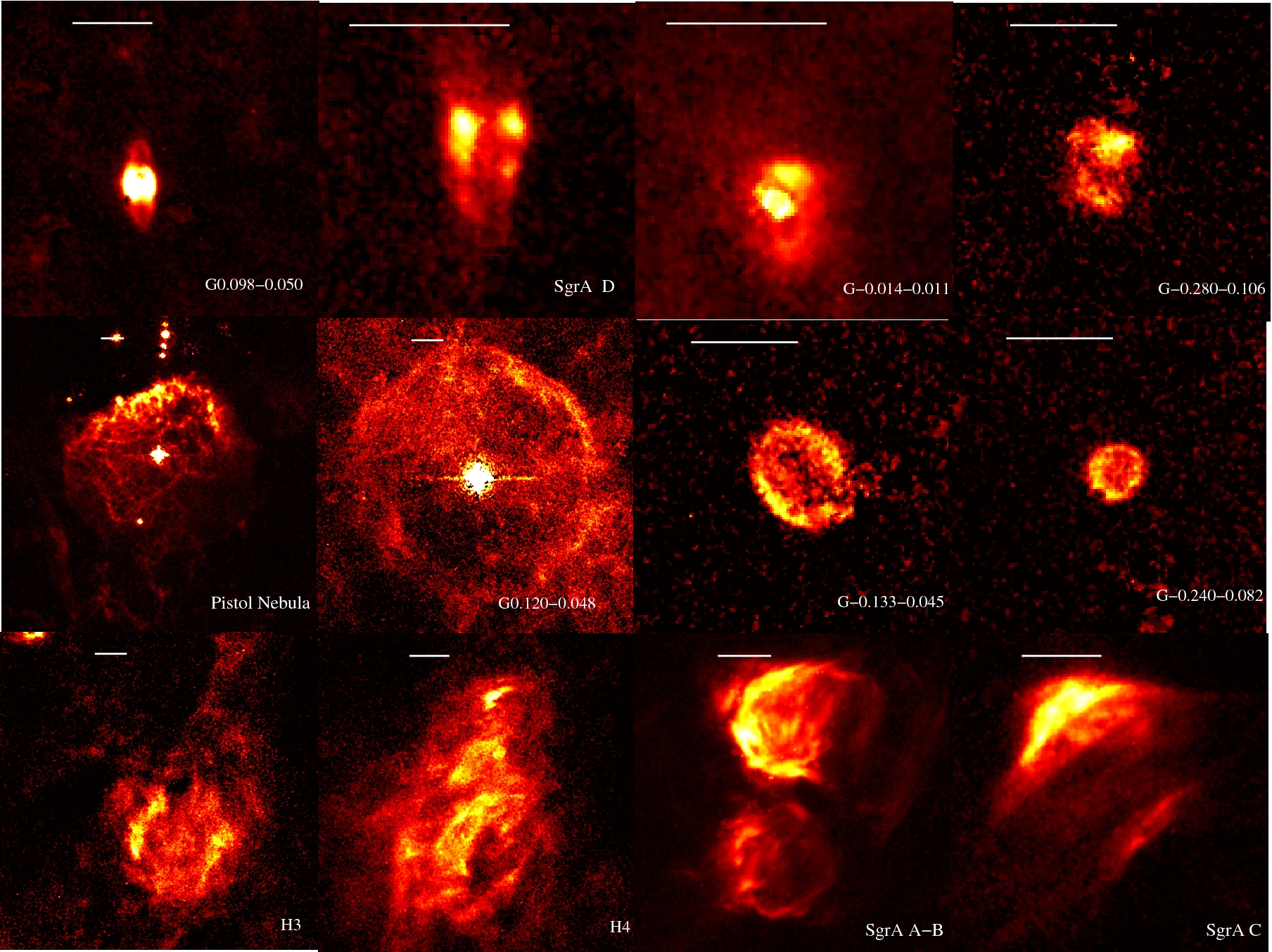}
%\centerline{\epsfig{figure=fig5.eps,width=1\textwidth,angle=0}}
\caption{%\footnotesize 
Close-ups of \pa\ images of selected individual 
nebulae (equatorially projected). Detected sources have been subtracted,
except for identified bright \pa\ emission star candidates, which help to 
show their potential relationship to the corresponding nebulae. 
The nebulae are labeled with rough Galactic 
coordinates, except for those with well-known names. 
The bar in each panel marks a 0.2 pc scale at the GC distance.}
\label{f:fig5}
\end{figure*}

Fig.~\ref{f:fig5} shows a selection of compact \pa-emitting 
nebulae.
%(CPNe). 
Some of these enigmatic 
nebulae may represent individual massive stars in formation (e.g., 
first row in Fig.~\ref{f:fig5}), 
whereas others may be due to stellar ejecta 
and/or their interaction with dense ambient medium (second and third rows).  
%The study of such objects can lead to better understanding of massive star formation and evolution processes. 

\subsection{Point-like sources}

The survey also enables us to perform an unprecedentedly uniform, high resolution, photometrically sensitive census of the stellar population. Our preliminary
detection based on the F190N continuum image gives 
a total of about 0.6 million 
%605,370
point-like sources, which accounts for about 86\% of the total 
observed F190N intensity in the survey field. 
This is the highest ratio of resolved point sources to continuum emission achieved to date in the very crowded GC field (e.g., Figs.~\ref{f:fig2}a and ~\ref{f:fig3}a). 
%Philipp et al. (1999) claimed a resolved fraction of 90% in K.
The median of the unresolved intensity
 is  $\sim 0.04 {\rm~mJy~arcsec^{-2}}$, averaged over the surveyed field.
Our 50\% source detection limit  is about 17th mag in the F190N band. A preliminary analysis shows that we should detect most of the massive main-sequence stars 
 (earlier than B3, corresponding to 
$\sim 8 M_\odot$) in the region
as well as the bulk of red giants, including
red clump stars. A detailed analysis of the spatial distribution 
of the detected sources and the still unresolved 
intensity will allow us to infer the overall intrinsic structure of
the stellar light distribution in the GC 
and to provide quantitative measurements of the foreground extinction.

Our preliminary analysis has resulted in the detection of about 200
\pa-emitting star candidates. Before the publication of such candidates,
however, we still need additional careful examination, especially 
for relatively weak sources, to minimize potential 
systematic contamination (e.g., due to local enhancements of diffuse 
\pa\ complexes). We easily detect all previously known emission line stars throughout the
survey region ($\sim 70$ total; mostly Wolf-Rayet stars). 
We have already verified an additional 30 emission line stars
via spectroscopic follow-up observations
~\citep[][Cotera et al. in preparation]{mau09}. We anticipate that
further follow up observations will at least double that number again.
More importantly, over half of the newly identified stars are located
outside the three known clusters. Therefore we have already more than doubled
the number of
known intercluster emission line stars.  The newly discovered stars will enable
us to address in greater detail the formation of massive stars outside
clusters in our upcoming papers. In particular, the \pa\ star candidates 
outside the clusters are not distributed
in a completely random fashion. 
%There is a strong concentration toward Sgr A East compact HII regions A-C, probably indicating a cluster in formation. 
In addition to the P$\alpha$ emission line, the F187N filter
contains a few helium lines which are often strong emission lines in
Wolf-Rayet stars.  Therefore the most easily detected stars in the F187N filter
are typically Wolf-Rayet stars, although future refinements to the data analysis
will enable us to detect O and possible B supergiants as well, which
have smaller \pa\ equivalent widths, in emission or absorption. 
Such refinements may be achieved by 
modeling the broad-band spectral energy distribution together with 
the narrow band measurements of the P$\alpha$ emission and absorption.
%Nevertheless, these initial results already indicate that a substantial number of massive stars are present outside the known clusters (Fig.~\ref{f:fig2}c). 

\section{Discussion}
\label{s:dis}

While the data analysis, 
multi-wavelength comparison, follow-up 
observations, and  modeling are still ongoing and will be presented in later papers, we discuss here the implications of the survey for addressing several
key issues related to 
star formation and its impact on the GC environment:

\subsection{How do massive stars form in the GC?}

Two large questions remain unanswered about massive star formation in
the GC: (1) Do massive stars form exclusively in the clusters,
or in looser associations throughout the field?  and (2) Is star
formation continuous, or does it happen in discrete bursts? Until recently,
massive (e.g., emission-line) stars have mostly be found
within the 3 GC clusters.
%(shown with circles in Fig.~\ref{f:fig2}a).  
A number of searches for young
massive stars at infrared and X-ray wavelengths have been carried out
~\citep[e.g., ][]{cot99,mun06a,mau07,mau09}.
These studies, each with diverse search criteria, have
yielded a rather incomplete picture of the young massive star population 
in this GC.

The P$\alpha$ emission is a sensitive
tracer of massive stars and their environs.
The emission traces warm ionized gas, produced and
maintained primarily by Lyman continuum radiation from massive stars. 
Such gas can be associated with individual stars (e.g., massive 
stellar winds), their immediate vicinities (circumstellar material),
and surrounding ISM structures. The fact that the majority of our new sources 
are actually located outside the three known clusters has strong implications for understanding
the dynamics of massive stars with ages of only a few $ \times 10^{6}$ yr. 
%The time scale for a cluster to dissolve in the GC remains quite uncertain (ranging from $\sim 10$ to 70 Myr, based on various N-body simulations 
% So we may expect to find a few similar clusters, if their formation rate has been constant over an extended period of time. We will systematically search for such clusters in our source distribution. 
The exact time scales for clusters to be
disrupted and the place where the stars end up provide
information about the internal dynamics of the clusters as well as
the large-scale tidal force of the GC~\citep[][]{kim99,kim00,kim04,kim03,por02,por03}. 
The spatial distribution of the massive stars detected
in our survey can be used to better constrain 
these processes throughout the field. The distributions of the  stars and 
their types as well as follow-up work on both radial
velocities and proper motions~\citep[][]{sto08} will allow us to 
constrain how quickly the massive stellar clusters are currently dissolving, 
how many star clusters have been 
disrupted, and whether or not the stars are formed in massive clusters which 
have been subsequently dispersed, or formed in relative isolation. 

%Based solely on the spatial distributions of stars as seen in both the broad continuum and Pa line observations, no obvious clusters, besides the three well known clusters have been identified to date. However, if there is a constant star formation rate, we would expect to see additional clusters in formation. Preliminary analysis suggests a few regions that have an overabundance of compact Pa sources, possibly indicative of clusters currently in formation process.   Dong et al. will explore in detail these interesting sources.  

\subsection{What are the state and dynamics of the ISM?}
In the high pressure and high density environment of the GC, 
the diffuse P$\alpha$-emitting gas tends to be found at ionization fronts 
bordering molecular clouds. The gas density and pressure are expected to
change drastically at such fronts~\citep[e.g., ][]{sco98}. The high
 resolution P$\alpha$ map resolves various ionization fronts, which
helps to determine the local neutral 
gas density and the thermal pressure of diffuse warm ionized gas. 
The comparison between Figs.~\ref{f:fig1} and \ref{f:fig2} demonstrates 
the complicated relationship among various
stellar and ISM components. Within the Sickle region
(Fig.~\ref{f:fig4}b), for instance, we see two distinctive ionization front morphologies.  We see columns of photo-evaporating gas similar to the elephant trunks or pillars seen in regions such as M16 in one part, but  also a remarkably smooth ionization front located on another side of the large HII region.  Although very different, both are morphologically perpendicular to the brightest non-thermal 
radio filaments, which are uniquely found in the GC field and have been 
a long-standing mystery. They are known to trace enhanced 
magnetic field and/or cosmic rays. But the underlying physical processes 
remain unclear. The observed relative morphological configuration then
suggests that the large-scale, inter-cloud, vertical (poloidal) magnetic 
field is combing through the dense gas and gets enhanced in the process. 
The expansion of the magnetized inter-cloud material, probably 
accompanied with cosmic ray acceleration, is likely driven by the strong 
mechanical energy output from the Quintuplet
and Arches clusters, consistent with the barrel-shaped morphology 
of the nonthermal radio filaments (Fig.~\ref{f:fig1}). 
Therefore, the feedback from the clusters 
likely plays an important role in shaping the ISM and the eco-system 
of the GC in general. 

\subsection{Comparison with other observations}
\label{ss:comparion}

A few P$\alpha$ observations were made previously in the GC region.  
The Sgr A West region, covering 
the inner $\sim$ 4 pc of the Galaxy, was observed using both NIC3 
and NIC2~\citep[e.g., ][]{sco03}. Several other programs 
concentrated only on the central parsec with NIC1. A NIC2 P$\alpha$ 
observation was also made for the core of the Arches cluster~\citep{fig02}. 
The P$\alpha$ emission from the 
Pistol Nebula (Fig.~\ref{f:fig4}b) was covered with a $2 \times 2$ mosaic
of NIC2 observations, which had a smaller field of view~\citep[19\farcs2 on a side; ][]{fig99}. 
Our map shows considerably more structure
in the diffuse \pa\ emission of the Nebula than the image shown 
in~\citet{fig99}, apparently due to the higher sensitivity of 
our NIC3 observations. 
In addition, our new data provide additional epochs for
variability analysis in these regions with the existing observations
~\citep[e.g., ][]{blu01}.

Our study of the GC region complements other efforts to understand
Galactic star forming regions, such as Orion~\citep[e.g., ][]{fei05}, 
the Carina nebula~\citep[e.g., ][]{tap06}, 
NGC 3603~\citep[][]{sto04,sto06}, and Westerlund 1~\citep[e.g., ][]{mun06b}. 
It is crucial to compare the time scales and efficiency 
with which stars form in the GC 
to those in other regions of the Galactic disk in order to gain a detailed 
understanding of the interplay between massive stars and the ISM. 

Ultimately, understanding the gas removal, via the
formation of massive stars and 
their impact on the ISM or other processes~\citep[e.g., Type Ia 
supernova-driven Galactic bulge wind; ][]{tan09,li09},  will help to determine why some 
host SMBHs accrete at rates near their Eddington limits, 
while others, including Sgr A*, are at rates orders of magnitude lower.
\section*{Acknowledgments}
We gratefully acknowledge the support of the staff at STScI 
for implementing the survey and for helping in the data reduction
and analysis. Support for program HST-GO-11120 was provided by NASA through a grant from the Space Telescope Science Institute, which is operated by the Association of Universities for Research in Astronomy, Inc., under NASA contract NAS 5-26555.
\label{lastpage}

\appendix


\begin{thebibliography}{99}
\bibitem[\protect\citeauthoryear{Arendt et al.}{2008}]{are08} Arendt R. G. et al., 2008, ApJ 682, 384
\bibitem[\protect\citeauthoryear{Blum et al.}{2001}]{blu01} Blum R. D. et al., 2001, AJ, 122, 1875
\bibitem[\protect\citeauthoryear{Chuss et al.}{2003}]{chu03} Chuss D. T. et al., 2003, ApJ, 599, 1116
\bibitem[\protect\citeauthoryear{Cotera et al.}{1996}]{cot96} Cotera A. S. et al.,  1996, ApJ, 461, 750
\bibitem[\protect\citeauthoryear{Cotera et al.}{1999}]{cot99} Cotera A. S. et al., 1999, ApJ, 510, 747 
\bibitem[\protect\citeauthoryear{Cotera et al.}{2006}]{cot06} Cotera A. et al., 2006, JPhCS, 54, 183
\bibitem[\protect\citeauthoryear{Cotera et al.}{2006}]{cot09} Cotera A. et al., 2009, in preparation
\bibitem[\protect\citeauthoryear{Diolaiti et al.}{2000}]{dio00} Diolaiti E. et al., 2000, , in Adaptive Optical Systems Technology, ed. P. L. Wizinowich, Proc. SPIE, 4007, 879
\bibitem[\protect\citeauthoryear{Dong et al.}{2009}]{don09} Dong H., et al., 2009, in preparation
\bibitem[\protect\citeauthoryear{Figer et al.}{1999}]{fig99} Figer D. F. et al., 1999, ApJ, 525, 759
\bibitem[\protect\citeauthoryear{Figer et al.}{2002}]{fig02} Figer D. F. et al.,  2002, ApJ, 581, 258
\bibitem[\protect\citeauthoryear{Feigelson et al.}{2005}]{fei05} Feigelson E. D. et al.,  2005, ApJS, 160, 379
\bibitem[\protect\citeauthoryear{Genzel et al.}{2003}]{gen03} Genzel R. et al., 2003, ApJ, 594, 812
\bibitem[\protect\citeauthoryear{Ghez et al.}{2008}]{ghe08} Ghez A. M. et al., 2008, ApJ, 689, 1044
\bibitem[\protect\citeauthoryear{Hummer \& Storey}{1987}]{hum87} Hummer D. G., \& Storey P. J., 1987, MNRAS 224,801
\bibitem[\protect\citeauthoryear{Kim, Morris \& Lee}{1999}]{kim99} Kim S. S., Morris M. \& Lee H. M., 1999, ApJ 525, 228
\bibitem[\protect\citeauthoryear{Kim et al.}{2000}]{kim00} Kim S. S., Figer D. F., Lee H. M., \& Morris M., 2000, ApJ 545, 301
\bibitem[\protect\citeauthoryear{Kim \& Morris}{2003}]{kim03} Kim S. S., \& Morris M., 2003, ApJ 597, 312
\bibitem[\protect\citeauthoryear{Kim, Figer \& Morris}{2004}]{kim04} Kim S. S., Figer D. F., \& Morris M., 2004, ApJ 607, L123
\bibitem[\protect\citeauthoryear{Lang et al.}{1997}]{lan97} Lang, C.~C., Goss W.~M., \& Wood D.~O.~S., 1997, ApJ, 474, 275
\bibitem[\protect\citeauthoryear{Lang et al.}{2001}]{lan01} Lang C.~C., Goss W.~M., \& Morris M., 2001, AJ, 121, 2681
\bibitem[\protect\citeauthoryear{Li, Wang \& Wakker}{2009}]{li09} Li Z. Y., Wang Q. D., \& Wakker B., 2009, MNRAS, 397, 148
\bibitem[\protect\citeauthoryear{Mauerhan, Muno \& Morris}{2007}]{mau07} Mauerhan J., Muno M., \& Morris M. R., 2007, ApJ, 662, 574
\bibitem[\protect\citeauthoryear{Mauerhan et al.}{2009}]{mau09} Mauerhan J. et al., 2009, ApJ, submitted
%Moffat, A. F. J. et al.\ 2002, ApJ, 573, 191\\
%Moneti, A., Stolovy, S. et al. 2001 A\&A 366, 106 \\
\bibitem[\protect\citeauthoryear{Morris}{1993}]{mor93} Morris M., 1993, ApJ, 408, 496
\bibitem[\protect\citeauthoryear{Morris \& Serabyn}{1996}]{mor96} Morris M., \& Serabyn E., 1996, ARA\&A, 34, 645
%Muno, M., et al. 2006a, ApJS, 165, 173\\
\bibitem[\protect\citeauthoryear{Muno et al.}{2006a}]{mun06a} Muno M. et al., 2006a, ApJ, 638, 183
\bibitem[\protect\citeauthoryear{Muno et al.}{2006b}]{mun06b} Muno M. P. et al., 2006b, ApJ, 650, 203
\bibitem[\protect\citeauthoryear{Muno et al.}{2009}]{mun09} Muno M. P. et al., 2009, ApJS, 181, 110
\bibitem[\protect\citeauthoryear{Nishiyama et al.}{2009}]{nis09} Nishiyama S. et al., 2009, ApJ., 690, 1648
\bibitem[\protect\citeauthoryear{Philipp et al.}{1999}]{phi99} Philipp S. et al., 1999, A\&A, 348, 768
\bibitem[\protect\citeauthoryear{Portegies Zwart et al.}{2002}]{por02} Portegies Zwart S. F. et al., 2002, ApJ, 565, 265
\bibitem[\protect\citeauthoryear{Portegies Zwart, McMillan \& Gerhard}{2003}]{por03} Portegies Zwart S. F., McMillan S. L. W., \& Gerhard O., 2003, ApJ, 593, 352
%Ramirez et al. 2008, ApJS, 175, 147\\
%Schneider, G. 2002, www.stsci.edu/hst/proposing/docs/\\$~\:\:\:$NICMOS-AO-whitepaper.html\\
%Sanchawala, K. et al. 2007, ApJ, 656, 462\\
\bibitem[\protect\citeauthoryear{Schultheis et al.}{2009}]{sch09} Schultheis M. et al., 2009, A\&A, 495, 157
\bibitem[\protect\citeauthoryear{Scoville et al.}{2003}]{sco03} Scoville N., et al., 2003, ApJ, 594, 294
\bibitem[\protect\citeauthoryear{Scowen et al.}{1998}]{sco98} Scowen et al., 1998, AJ, 116, 163 
\bibitem[\protect\citeauthoryear{Serabyn, Shupe \& Figer}{1998}]{ser98} Serabyn E., Shupe D., \& Figer D. F., 1998, Nature, 394, 448
\bibitem[\protect\citeauthoryear{Simpson et al.}{1997}]{sim97} Simpson, J. P., et al., 1997, ApJ, 487, 689
\bibitem[\protect\citeauthoryear{Smith, Stassun \& Bally}{2005}]{smi05} Smith N. Stassun K. G., \& Bally J., 2005, ApJ, 129, 888
\bibitem[\protect\citeauthoryear{Stolovy et al.}{2006}]{stolo06} Stolovy S. et al., 2006, JPhCS, 54, 176
\bibitem[\protect\citeauthoryear{Stolte et al.}{2004}]{sto04} Stolte A. et al., 2004, AJ, 128, 765
\bibitem[\protect\citeauthoryear{Stolte et al.}{2006}]{sto06} Stolte A. et al., 2006, AJ, 132, 253
\bibitem[\protect\citeauthoryear{Stolte et al.}{2008}]{sto08} Stolte A. et al., 2008, ApJ, 675, 1278
\bibitem[\protect\citeauthoryear{Tang et al.}{2009}]{tan09} Tang S. K. et al., 2009, MNRAS, 392, 77
\bibitem[\protect\citeauthoryear{Tapia et al.}{2006}]{tap06} Tapia M. et al., 2006, MNRAS, 367, 513
\bibitem[\protect\citeauthoryear{Wang, Gotthelf \& Lang}{2002}]{wan02} Wang Q. D., Gotthelf E., \& Lang C.~C., 2002, Nature, 415, 148
\bibitem[\protect\citeauthoryear{Wang, Dong \& Lang}{2006}]{wan06} Wang Q. D., Dong H, \& Lang C.~C.,  2006, MNRAS, 371, 38
\bibitem[\protect\citeauthoryear{Yusef-Zadeh, Morris \& Chance}{1984}]{yus84} Yusef-Zadeh F., Morris M., \& Chance D., 1984, Nature, 310, 557
%Yusef-Zadeh, F. \& Morris, M. 1987, ApJ, 322, 721\\
\bibitem[\protect\citeauthoryear{Yusef-Zadeh et al.}{2002}]{yus02} Yusef-Zadeh F. et al., 2002, ApJ, 570, 665

\end{thebibliography}
\end{document}